\pdfoutput=1
\documentclass[11pt]{article}
\usepackage[letterpaper,margin=1in]{geometry}
    
\usepackage{amsmath,amssymb,amsthm,booktabs,microtype,xcolor}
\usepackage[numbers,sort&compress]{natbib}
\usepackage{hyperref}
\hypersetup{colorlinks=true,linkcolor=blue!50!black,citecolor=blue!50!black,urlcolor=blue!50!black}
\usepackage{doi}
\usepackage{algorithm}
\usepackage{algpseudocode}
\usepackage{tabularx}
\usepackage{graphicx}
\newtheorem{definition}{Definition}
\newtheorem{theorem}{Theorem}
\newtheorem{proposition}{Proposition}
\newtheorem{lemma}{Lemma}
\newtheorem{example}{Example}
\newcommand{\AP}{\mathsf{AP}}
\newcommand{\Target}{\varphi_T}
\newcommand{\Hyp}{\varphi_H}

\newcommand{\Hints}{\mathcal H_{\mathrm{hint}}}

\newcommand{\mq}{\mathsf{MQ}}

\title{Active Learning of Temporal Specifications \\
through Distinguishing Queries}
\author{Rajarshi Roy \qquad Xiaowei Huang\\[2pt] University of Liverpool}
\date{}
\begin{document}
\maketitle

\begin{abstract}
Temporal specifications are essential for verification and monitoring,
yet writing formulas that capture the intended behaviour is difficult.
Active learning derives such specifications through interaction between
a learner and a teacher, but remains underexplored for temporal logics.
Existing approaches require the teacher to answer equivalence queries
and to construct counterexamples, and they often rely on SAT-based
learning procedures that restart the search after each new example. We present
an active-learning framework for LTLf in which the teacher only labels
traces and answers natural-language questions about the intended
behaviour. The learner maintains an enumerated version space of
candidate formulas, generates traces that distinguish candidates, and
uses the teacher's hints to rank candidates or to
restrict the search. We prove consistency with the observed examples,
identification within bounded hypothesis classes with a bound on the
number of queries, and a PAC guarantee for the final hypothesis. In
experiments with a simulated teacher on $36$ targets, including targets
in four standard environments, the learner identifies every target when
it compares all rivals, with fewer than seven queries beyond an initial sample of ten traces on
average. Correct hints reduce the number of queries by up to $16\%$ and
the number of generated candidates by a fifth, and hints used for
ranking never prevent identification, even when incorrect. Our
enumeration-based version space is faster than the SAT-based approach
of previous learners at every target size.
\end{abstract}

\section{Introduction}
Temporal logics are ubiquitous in formal methods, providing a compact and
interpretable language for expressing system requirements in domains including
verification~\cite{pnueli1977temporal,vardi1986automata}, runtime
monitoring~\cite{bauer2011runtime}, and
planning~\cite{bacchus2000using}.
Recently, they have been employed in the verification of machine learning
systems~\cite{seshia2022toward,hasanbeig2023certified}, agentic
frameworks~\cite{yang2024plug,kamath2025enforcing}, and robotic
systems~\cite{kressgazit2021formalizing}.

However, formalising the intended behaviour in a temporal logic can be a complex and error-prone process, in particular for non-experts~\cite{DBLP:conf/secdev/DansoHBC26,DBLP:journals/programming/GreenmanSNK23}.
Active learning provides a framework for identifying correct specifications through interaction with a teacher who is assumed to know the target specification.
In a typical exact-learning loop, the learner poses membership queries,
asking whether a specific execution satisfies the target specification,
and equivalence queries, asking whether the current hypothesis is
equivalent to the target. For a failed equivalence query, the teacher
provides a counterexample on which the hypothesis and target disagree.
The learner uses this feedback to refine its
hypothesis~\cite{angluin1987learning,angluin1988queries,vaandrager2017model}.

Equivalence queries and counterexample construction demand formal
expertise that non-expert users may lack. Such users may nevertheless
judge whether an execution is acceptable and explain their concerns in
natural language, motivating learning from execution labels and informal
guidance.

Existing active learners for temporal logics share a second limitation.
They encode the search for a consistent formula as a SAT or
optimisation problem~\cite{camacho2019learning,gavran2020interactive},
following SAT-based passive learning~\cite{neider2018learning}. Such a
learner either solves the problem anew after each
example~\cite{camacho2019learning}, or computes a small set of
candidates once and does not refine them as new examples arrive~\cite{gavran2020interactive}.
Since a formula has many encodings, every solution must be blocked
encoding by encoding, and the search becomes slow as the class of
formulas grows.

We present a modular active-learning framework for LTLf in which the
teacher only labels traces and, when able, answers natural-language
questions. The learner reuses its search by maintaining a \emph{version
space} of enumerated formulas consistent with the observed labels.

To refine its hypothesis, the learner selects a rival from the version
space and asks the teacher to label a trace on which the two disagree.
Each label eliminates all inconsistent candidates, allowing efficient
learning even from millions of formulas. If no distinguishing trace
exists, the rival is equivalent to the hypothesis and is discarded
without a query. Our instantiation constructs the version space using
grammar-based enumeration~\cite{DBLP:conf/ijcai/0002PPK25} and generates
distinguishing traces using automata~\cite{degiacomo2013linear} or formula
progression~\cite{bacchus2000using}. SAT-based approaches can also be
incorporated. Traces may be \emph{free}, allowing arbitrary event
sequences, or constrained by an \emph{environment}, ensuring that the
teacher only labels feasible executions.

Membership labels indicate whether an execution is acceptable, but not why. To provide this additional information, our framework introduces \emph{natural-language queries} at two levels. At Level~1, the teacher identifies the type of mismatch, such as a missing event or incorrect ordering. At Level~2, the teacher specifies the requirement more precisely, identifying the events involved and their intended relationship. Both levels guide the learner in restricting the candidate space, potentially reducing the number of queries needed.

\begin{example}[Motivating example]
\label{ex:motivating}
To illustrate the framework, consider a mobile service robot that picks up objects and delivers them.
The user knows the intended behaviour but cannot write it in temporal
logic. Over $\AP=\{\mathit{pickup},\mathit{deliver}\}$, the target is
\(
\Target=\mathbf G\bigl(\mathit{pickup}\rightarrow
\mathbf X(\neg\mathit{pickup}\,\mathbf U\,\mathit{deliver})\bigr):
\)
every pickup is followed later by a delivery, with no other pickup in
between. We follow a run of our learner with a simulated teacher,
started from two labelled traces.

First, the user labels two executions: $\{\mathit{pickup}\};\{\mathit{deliver}\}$
is accepted and $\{\mathit{pickup},\mathit{deliver}\}$ is rejected. Asked
whether something required is missing, the user answers that the object
should be delivered after a pickup; this hint becomes the requirement
$h=\mathbf G(\mathit{pickup}\rightarrow\mathbf F\,\mathit{deliver})$, and
the learner first searches the formulas consistent with it. Then, the
learner repeatedly compares its hypothesis with other consistent
formulas and asks the user about an execution on which they disagree.
For instance, when the current hypothesis is
$\mathbf G(\mathit{pickup}\rightarrow\mathbf F\mathbf X\,\mathit{deliver})$
and the rival is $\Target$, the user rejects the distinguishing trace
$\{\mathit{pickup}\};\{\mathit{pickup}\};\{\mathit{deliver}\}$.
This refutes the hypothesis, so the rival $\Target$ becomes the new
hypothesis. Finally, when no remaining
formula can be told apart from the hypothesis, the learner returns
$\Target$, after $16$ membership queries including the two initial ones;
without the hint, it needs $20$.
\end{example}

Our contributions therefore are:
\begin{itemize}
    \item a modular active-learning framework for LTLf that uses membership
    queries and natural-language hints
    (Section~\ref{sec:approach});
    \item different levels of natural-language hints to focus on relevant formulas only (Section~\ref{sec:nl-hints});
    \item guarantees of sample consistency, identification within
    bounded hypothesis classes with a bound on the number of queries,
    and a PAC guarantee for the final hypothesis
    (Section~\ref{sec:properties});
    \item an open-source modular Python implementation with replaceable
    procedures for version-space construction and distinguishing-trace
    generation (Section~\ref{sec:implementation}), evaluated over
    $5{,}540$ runs on $36$ LTLf targets expressing specification patterns
    used in practice~\cite{dwyer1999patterns}, including $16$ targets
    across four standard environments (Section~\ref{sec:experiments});

\item experimental evidence that the learner identifies all targets when comparing all rivals, requiring on average only $6.2$ additional queries over free traces and $2.5$ in environments, even with up to $5.1$ million candidate formulas. Correct hints reduce queries by up to $16\%$, while our version-space construction outperforms the SAT-based alternative ($8.8$\,s versus $22.4$\,s for size-$7$ targets).
\end{itemize}
Proofs and further technical and experimental details are given in
the appendices.

\subsection{Related work}
Active learning has its roots in automata learning, notably Angluin's $L^*$ algorithm~\cite{angluin1987learning}, which has been widely applied to learning behavioural models of software and hardware systems~\cite{vaandrager2017model}. Automata learning benefits from the Myhill--Nerode relation, which connects observed behaviour to the structure of the learned model. This connection is less direct for logical formalisms, including temporal logics, motivating separate approaches to actively learning logical specifications.

Three works on active learning of temporal specifications are particularly close to ours. Camacho and McIlraith~\cite{camacho2019learning} learn minimal LTLf formulas using equivalence queries. Their approach repeatedly invokes a SAT-based passive learner to find the smallest formula consistent with the accumulated examples, incorporating counterexamples returned by the oracle after each failed equivalence query. Linard and T{\r u}mov{\'a}~\cite{linard2020active} learn STL specifications using membership and equivalence queries. Their learner maintains an online decision tree, generates signals for membership queries, and evaluates equivalence through a distance to the target, obtaining labelled counterexamples when equivalence fails. Gavran et al.~\cite{gavran2020interactive} synthesise LTL specifications for robotic tasks from an example trace and a natural-language description. They use an optimisation-modulo-theories encoding to generate candidate specifications and ask users to assess behaviours that distinguish them.

Our framework differs from these works in three main respects. First, the teacher only labels traces and answers natural-language questions. Unlike in~\cite{camacho2019learning,linard2020active}, the teacher need not evaluate hypotheses or construct counterexamples. Second, the learner maintains all consistent formulas within a bounded class, guaranteeing identification of the target within a bounded number of queries (Section~\ref{sec:properties}). In contrast, the restricted candidate set of~\cite{gavran2020interactive} may exclude the target, while convergence in~\cite{linard2020active} is demonstrated experimentally but not formally established. Third, our version space is constructed once and incrementally filtered as new labels arrive, whereas~\cite{camacho2019learning} solves a new SAT problem after each counterexample. Our experiments show that enumeration is faster than SAT-based construction across all evaluated target sizes.

Natural language also serves a different purpose. While~\cite{gavran2020interactive} uses an initial natural-language description of the task, our framework elicits partial requirements through optional natural-language queries during learning, complementing membership labels. Finally, the framework is modular, allowing individual procedures to be replaced. For instance, a SAT-based version space can reduce memory consumption for larger hypothesis classes (Section~\ref{sec:rq-cost}).

A closely related area is \emph{passive learning}, which infers formulas from fixed datasets and often serves as a subroutine in active learning. For LTL and LTLf, notable approaches include SAT-based inference~\cite{neider2018learning}, scalable anytime learning~\cite{raha2022scalable}, graph-neural-network-based inference~\cite{luo2022bridging}, learning from positive examples~\cite{roy2023learning}, statistical inference from event logs~\cite{chiariello2024learning}, and grammar-based enumeration with search heuristics~\cite{DBLP:conf/ijcai/0002PPK25}, which forms the basis of our version-space construction. For STL, existing methods include decision-tree learning~\cite{bombara2021offline}, neural-symbolic approaches~\cite{li2023learning}, and optimisation-based decision-tree learning~\cite{liang2024learning}. More recently, Baharisangari et al.~\cite{baharisangari2025logic} have applied STL inference to causal discovery from time-series data.

\section{Preliminaries}

\subsection{Linear Temporal Logic on Finite Traces}
To illustrate the framework, we use linear temporal logic on finite traces
(LTLf) and introduce the necessary definitions~\cite{degiacomo2013linear}.
We represent relevant system properties by a finite set of atomic
propositions $\AP$. The alphabet $\Sigma=2^{\AP}$ consists of all subsets
of $\AP$; each letter specifies which propositions hold at a given
position. The trace domain $\mathcal W$ is $\Sigma^*$, which includes the empty
trace $\varepsilon$, unless an environment restricts it to feasible
traces (Section~\ref{sec:properties}).

The syntax of LTLf formulas is defined by the grammar
\[
\varphi ::= p \in\AP \mid \neg\varphi
\mid \varphi \land \varphi
\mid \mathbf X\varphi
\mid \varphi\,\mathbf U\,\varphi.
\]
We also use the standard derived Boolean
operators $\lor$ and $\rightarrow$, constants $\mathrm{false}$ and $\mathrm{true}$,
and the temporal operators $\mathbf F$ (eventually) and
$\mathbf G$ (always).
Although our framework supports formulas of arbitrary syntactic structure, we restrict negation to atomic propositions, following the equivalent negation normal form used in~\cite{DBLP:conf/ijcai/0002PPK25}, to simplify the search.

For a finite trace $w$ and an index $0\leq i\leq |w|$, satisfaction
is defined as follows:
\[
\begin{aligned}
w,i\models p &\quad\text{iff }i<|w|\text{ and }p\in w_i,\\
w,i\models\neg\varphi &\quad\text{iff }w,i\not\models\varphi,\\
w,i\models\varphi\land\psi &\quad\text{iff }w,i\models\varphi
  \text{ and }w,i\models\psi,\\
w,i\models\mathbf X\varphi &\quad\text{iff }i+1<|w|\text{ and }\\
&\qquad w,i+1\models\varphi,\\
w,i\models\varphi\,\mathbf U\,\psi &\quad\text{iff there is }j\geq i
  \text{ with }j<|w|,\ w,j\models\psi,\\
&\qquad \text{ and }w,k\models\varphi \text{ for every }i\leq k<j.
\end{aligned}
\]
We write $w\models\varphi$ for $w,0\models\varphi$ and
$L(\varphi)=\{w\in\mathcal W\mid w\models\varphi\}$.
We define equivalence and implication through the corresponding languages:
$\varphi\equiv\psi$ means $L(\varphi)=L(\psi)$, and
$\varphi\models\psi$ means $L(\varphi)\subseteq L(\psi)$.

Following \cite{DBLP:conf/ijcai/0002PPK25}, we measure formula
complexity by size and temporal depth. The size $|\varphi|$ is
the number of syntax-tree nodes, counting each literal (including
negated propositions) as a single leaf. The temporal depth
$\operatorname{td}(\varphi)$ is the maximum number of temporal operators
along any root-to-leaf path. For example,
$\mathbf G(p\rightarrow\mathbf F(\neg q))$
has size $5$ and temporal depth $2$.

\subsection{Background on Active Learning}
Active learning~\cite{angluin1987learning} involves two entities: a learner and a teacher.
The teacher represents a non-expert user who knows the intended target specification $\Target$ and responds to the learner's queries. 
The learner interacts with the teacher through two types of queries, formalised below: membership queries, which request the teacher to classify traces according to the target specification, and natural-language queries, which elicit hints to guide the refinement of the learned specification.
Note that, in contrast to traditional active learning, our framework avoids equivalence queries, as non-expert users cannot be expected to assess whether a learned specification precisely captures their intentions.

\begin{definition}[Membership query]
Given a target $\Target$ and a trace $w\in\mathcal W$, a membership
query $\mq(w)$ returns $1$ if $w\models\Target$ and $0$ otherwise.
\end{definition}

\begin{definition}[Natural-language query]
Given a target $\Target$, a natural-language query returns a hint $h$
describing a property of $\Target$.
\end{definition}

The traces labelled so far and their labels form a labelled \emph{sample}
$S\subseteq\mathcal W\times\{0,1\}$. A formula $\varphi$ is \emph{consistent}
with $S$ if, for every $(w,y)\in S$, $w\models\varphi$ if and only if $y=1$.

\section{Learning Problem}
Fix a finite set of propositions $\AP$ and the trace domain
$\mathcal W\subseteq(2^{\AP})^*$. Let $\mathcal L_{n,d}$ denote the finite
class of LTLf formulas over $\AP$ with size at most $n$ and temporal
depth at most $d$ (Section~\ref{sec:version-space}).
The teacher answers membership queries according to an unknown target
$\Target\in\mathcal L_{N,D}$, where $N,D$ are unknown to the learner.

The objective is to identify a hypothesis $\Hyp$ using as few
membership queries as possible, optionally guided by natural-language
hints. Every hypothesis must be consistent with the labelled sample
$S=\{(w,\mq(w))\mid w\text{ was queried}\}$.
We distinguish two objectives:
(i) \emph{exact identification}, $\Hyp\equiv\Target$, when the
searched class contains $\Target$ and refinement is exhaustive;
and (ii) \emph{approximate identification} under a distribution
$\mathcal D$ over $\mathcal W$, measured by
\[
\operatorname{err}_{\mathcal D}(\Hyp,\Target)
=\Pr_{w\sim\mathcal D}
[\mathbf1[w\models\Hyp]\ne\mathbf1[w\models\Target]].
\]
For $\epsilon,\delta\in(0,1)$, certification must satisfy
\[
\Pr[\text{accept }\Hyp\ \land\
\operatorname{err}_{\mathcal D}(\Hyp,\Target)>\epsilon]
\leq\delta.
\]
This guarantee does not imply exact equivalence.

\section{Active Learning Framework}
\label{sec:approach}

The learner starts from an initial hypothesis and refines it
iteratively, using several interchangeable procedures
(Algorithm~\ref{alg:main}).

\begin{algorithm}[t]
\caption{Active learning by distinguishing queries}
\label{alg:main}
\begin{algorithmic}[1]
\Require Class schedule $\mathcal L_1\subseteq\cdots\subseteq\mathcal L_M$,
  shortlist size $K\in\mathbb N\cup\{\infty\}$, budgets $L,P$
\Ensure Hypothesis $\Hyp$, labelled sample $S$,
  certification result $b$ and counterexample $w$
\State $S\gets\textsc{Initialize}(L)$
\State $\Hints\gets\emptyset$; $\Hyp\gets\bot$; $V\gets\emptyset$;
  $\mathcal L_0\gets\emptyset$ \Comment{$\bot$: no hypothesis yet}
\For{$m=1,\ldots,M$}
  \State $V\gets V\cup\textsc{VersionSpace}(\mathcal L_m\setminus
    \mathcal L_{m-1},S)$
  \State \textbf{if} $V=\emptyset$ \textbf{then continue}
  \State \textbf{if} $\Hyp=\bot$ \textbf{then} $\Hyp\gets$ a smallest
    formula in $V$
  \State $\Hints\gets\Hints\cup\textsc{AskHint}(\Hyp)$
    \Comment{at most once}
  \Repeat
    \State $R\gets$ first $K$ formulas of $\textsc{NextRival}(V,\Hyp,\Hints)$
    \ForAll{$\psi\in R$}
      \State $w\gets\textsc{Distinguish}(\Hyp,\psi)$
      \If{$w=\mathrm{false}$}
        \State $V\gets V\setminus\{\psi\}$
          \Comment{$\psi\equiv\Hyp$}
      \Else
        \State $S\gets S\cup\{(w,\mq(w))\}$
        \State \textbf{if} $\Hyp$ is inconsistent with $S$
          \textbf{then} $\Hyp\gets\psi$
        \State $V\gets\{\varphi\in V\mid\varphi
          \text{ is consistent with } S\}$
        \State \textbf{break}
          \Comment{Start the next round}
      \EndIf
    \EndFor
  \Until{no query was asked in this round}
\EndFor
\State \textbf{if} $\Hyp=\bot$ \textbf{then return}
  $(\textsc{Failure},S)$
\State $(b,w)\gets\textsc{Certify}(\Hyp,P)$
\State \Return $(\Hyp,S,b,w)$
\end{algorithmic}
\end{algorithm}

$\textsc{Initialize}(L)$ labels $L$ initial traces (line~1), sampled uniformly, generated as feasible environment rollouts, or provided as input. The learner then searches a \emph{class schedule}, an increasing sequence of finite classes $\mathcal L_1\subseteq\cdots\subseteq\mathcal L_M$ that fixes the order in which formulas are considered; by default, $\mathcal L_m=\mathcal L_{m,d}$ for a fixed depth $d$, so smaller formulas come first. For each class, \textsc{VersionSpace} adds formulas consistent with the labelled sample $S$ to the version space $V$ (line~4; Section~\ref{sec:version-space}), and the smallest formula in $V$ becomes the initial hypothesis (line~6). \textsc{AskHint} may request natural-language hints to reorder rivals or adjust the class schedule (line~7; Section~\ref{sec:nl-hints}).

In each refinement round (lines~8--21), \textsc{NextRival} orders the remaining candidates, called \emph{rivals}, and the learner considers the first $K$ (Section~\ref{sec:next-rival}). If \textsc{Distinguish} finds no distinguishing trace, the rival is equivalent to $\Hyp$ and is discarded without a query (line~13; Section~\ref{sec:distinguish}). Otherwise, the teacher labels the returned trace $w$ (line~15). If the label refutes $\Hyp$, the rival becomes the new hypothesis (line~16), and all inconsistent formulas are removed from $V$ (line~17). A round without a query ends refinement for the current class. Finally, \textsc{Certify} evaluates the hypothesis on $P$ fresh traces (line~24; Section~\ref{sec:certify}). The implementation also limits the number of comparisons (Section~\ref{sec:implementation}).

\paragraph*{Correctness and cost.}
Each query refutes either $\Hyp$ or the compared rival, eliminating at least one candidate without removing the target. Thus, the hypothesis remains consistent with all observed labels (Proposition~\ref{prop:sample-consistency}). When all rivals are compared ($K=\infty$, the \emph{exhaustive rule}), the learner identifies a hypothesis equivalent to the target (Theorem~\ref{thm:bounded-identification}), or equivalent relative to an environment (Theorem~\ref{thm:env-identification}). Comparing only the first $K$ rivals (the \emph{top-$K$ rule}) does not guarantee equivalence, but the certification bound (Theorem~\ref{thm:pac}) still applies. Under the exhaustive rule, the choice of version space and rival ordering affects only query efficiency and running time.

We now describe each procedure in turn.

\subsection{Version space}
\label{sec:version-space}

For a class $\mathcal L$ and a sample $S$, the version space is
\[
V_{\mathcal L}(S)=\{\psi\in\mathcal L\mid
w\models\psi\text{ iff }y=1\text{ for all }(w,y)\in S\}.
\]
$\textsc{VersionSpace}(\mathcal L,S)$ returns $V_{\mathcal L}(S)$.
Algorithm~\ref{alg:main} calls it only for the formulas that are new in
the schedule (line~4), and it keeps the version space up to date after
each label (line~17). We describe two implementations.

\paragraph{Enumeration.}
Following the grammar-based enumeration of Roy et
al.~\cite{DBLP:conf/ijcai/0002PPK25}, in the first approach, we build the version space
bottom-up by dynamic programming. Starting from the atomic propositions
and their negations, the formulas of each size and nesting depth are
built from smaller ones with the operators $\mathbf X$, $\mathbf F$,
$\mathbf G$, $\mathbf U$, $\land$, $\lor$ and $\rightarrow$. A new
formula enters the version space only if it is consistent with the current
$S$.

As in~\cite{DBLP:conf/ijcai/0002PPK25}, two heuristics discard
redundant formulas. \emph{Temporal simplification} discards the forms
$\mathbf F\mathbf F\varphi$, $\mathbf G\mathbf G\varphi$,
$\mathbf X\varphi\,\mathbf U\,\mathbf X\psi$,
$\mathbf G\varphi\land\mathbf G\psi$, $\mathbf X\varphi\land\mathbf X\psi$,
$\mathbf F\varphi\lor\mathbf F\psi$ and $\mathbf X\varphi\lor\mathbf X\psi$,
each equivalent over finite traces to a smaller formula. \emph{Boolean
simplification} discards binary formulas whose operands are
syntactically equal or complementary literals. Since larger formulas are
built from smaller ones, all formulas of size below $n$ are stored,
consistent or not, so time and memory grow with $|\mathcal L_{n,d}|$.

\paragraph{SAT-based construction.}
Our second approach follows SAT-based passive
learning~\cite{neider2018learning,camacho2019learning} and represents
the version space implicitly by a propositional formula
$\Phi^S_{m,d}=\phi_{\mathrm{syn}}\land\phi_{\mathrm{sem}}$, whose
satisfying assignments encode the formulas of size $m$ and temporal
depth at most $d$ that are consistent with $S$: $\phi_{\mathrm{syn}}$
encodes their syntax and $\phi_{\mathrm{sem}}$ their semantics on the
sample traces.

For $\phi_{\mathrm{syn}}$, a formula of size $m$ is represented by its
syntax tree with nodes $1,\ldots,m$: the root is node $m$, children have
smaller indices than their parents, leaves are propositions or their
negations, and inner nodes are operators. The variable $x_{i,\lambda}$
indicates that node $i$ is labelled by the literal or operator
$\lambda$, and $l_{i,j}$ and $r_{i,j}$ indicate that node $j$ is the left
or right child of node $i$. The constraints assign exactly one label to
each node and give each operator as many children as its arity. Unlike the syntax DAGs
of~\cite{neider2018learning}, every non-root node has exactly one
parent, so the encoding represents syntax trees and matches our size
measure. In addition, the variable $e_{i,k}$ indicates that the subformula
rooted at node $i$ has temporal depth at most $k$, and the constraints
propagate these values from the leaves to the root, so that the root has temporal depth at most $d$.

For $\phi_{\mathrm{sem}}$, each trace $w$ of $S$ is evaluated position by
position. The variable $z^w_{i,k}$ indicates that the subformula rooted
at node $i$ holds at position $k$ of $w$, for $0\leq k\leq|w|$, where
$k=|w|$ is the end position. The constraints relate these variables to
the labels and children of the nodes, following the finite-trace
semantics of the preliminaries, and require $z^w_{m,0}=y$ for each
$(w,y)\in S$. Thus $\Phi^S_{m,d}$ is satisfiable if and only if some
formula of size $m$ and temporal depth at most $d$ is consistent with
$S$, and such a formula can be decoded from any satisfying assignment.

In this approach, the version space need not be listed explicitly. 
\textsc{NextRival} asks an incremental SAT solver for models of $\Phi^S_{m,d}$, in increasing order
of $m$. If \textsc{Distinguish} proves the decoded rival $\psi$
equivalent to $\Hyp$, a blocking clause, as
in~\cite{neider2018learning,gavran2020interactive}, excludes the current assignment
to the structural variables $x_{i,\lambda}$, $l_{i,j}$ and $r_{i,j}$.

\subsection{Next rival}
\label{sec:next-rival}

$\textsc{NextRival}(V,\Hyp,\Hints)$ lists the formulas of
$V\setminus\{\Hyp\}$ in some order, and Algorithm~\ref{alg:main}
compares $\Hyp$ with the first $K$ of them. The order decides which
formulas are compared, and hence which traces are queried. We consider
three orders.

\paragraph{Enumeration.}
The rivals come in the order in which the version space produces them:
smallest formulas first, with ties broken by a fixed textual order. With
the SAT-based construction, this is the order in which the solver
returns models, again smallest size first. The order needs no extra
computation.

\paragraph{Semantic.}
The rivals closest to $\Hyp$ in behaviour come first. For a probe set
$U=\{u_1,\dots,u_r\}$, the semantic distance
\[
d_{\rm sem}(\varphi,\psi;U)=\frac1r\sum_{i=1}^r
\mathbf1[\mathbf1[u_i\models\varphi]\ne\mathbf1[u_i\models\psi]]
\]
is the fraction of probe traces on which the two formulas disagree. The
probe set is drawn once per run, independently of the sample, from the
distribution of the initial traces, or as feasible rollouts in an
environment, so $d_{\rm sem}$ estimates the probability that the two
formulas disagree on such a trace. We also evaluate the reverse order,
\emph{semantic-far}, which puts the farthest rivals first.

\paragraph{Split.}
We prioritise rivals whose witnesses split the version space as evenly as possible. 
A more balanced split guarantees that more candidates are eliminated, regardless of how the teacher labels the trace.

For each rival $\psi$, let $w_\psi$ be the first trace in breadth-first order in the short-trace pool $T_\ell$ (Section~\ref{sec:distinguish}) on which $\psi$ and $\Hyp$ disagree, if such a trace exists. Let $n_\psi=|\{\varphi\in V\mid w_\psi\models\varphi\}|$ count the candidates that accept $w_\psi$. We rank rivals by increasing imbalance $\bigl|2n_\psi-|V|\bigr|$. Rivals with no witness in $T_\ell$ come last, in enumeration order.

\subsection{Distinguish}
\label{sec:distinguish}

$\textsc{Distinguish}(\varphi,\psi)$ returns a trace in $\mathcal W$ on which $\varphi$ and $\psi$ disagree, or $\mathrm{false}$ if no such trace exists. One witness suffices per comparison because either label makes one of the two formulas inconsistent with the sample.

Before running either of the two sound and complete procedures below, we scan a fixed pool $T_\ell$ containing all traces of length at most $\ell$, restricted to feasible traces when an environment is given. We scan the pool in breadth-first order and return the first trace on which the formulas disagree. If the pool contains no witness, we run the chosen procedure.

\subsubsection{Distinguishing with automata}
The first procedure translates each LTLf formula $\varphi$ into a
deterministic finite automaton (DFA) $\mathcal A_\varphi$ with the same
language~\cite{degiacomo2013linear}. For candidates $\varphi$ and
$\psi$, the product construction~\cite{rabin1959finite}, in which a state
is accepting exactly when one of its components is, yields a DFA
$\mathcal A_\Delta$ with
$L(\mathcal A_\Delta)=L(\mathcal A_\varphi)\mathbin{\triangle}L(\mathcal A_\psi)$.
Every trace of $L(\mathcal A_\Delta)$ is accepted by exactly one
candidate, so the procedure returns such a trace if $\mathcal A_\Delta$ is
nonempty, and $\mathrm{false}$ otherwise.

\subsubsection{Distinguishing by formula progression}
\label{sec:distinguish-prog}
The automata-based procedure builds an automaton for every compared
formula. Our implementation instead explores the symmetric-difference
automaton on the fly, using formula progression~\cite{bacchus2000using}.
For a letter $a$, the progression $\operatorname{prog}(\varphi,a)$ is the
formula that the rest of the trace must satisfy once $a$ has been read; it
extends to traces letter by letter. The \emph{end value}
$\operatorname{fin}(\varphi)\in\{0,1\}$ states whether $\varphi$ holds on
the empty trace. Both are computed syntactically, and $uw\models\varphi$
if and only if $w\models\operatorname{prog}(\varphi,u)$.
Appendix~\ref{app:progression} gives the definitions.

\begin{example}
Let $\AP=\{a,b\}$ and $\varphi=a\,\mathbf U\,b$. Then
$\operatorname{prog}(\varphi,\{a\})=\varphi$,
$\operatorname{prog}(\varphi,\{b\})=\mathrm{true}$ and
$\operatorname{prog}(\varphi,\emptyset)=\mathrm{false}$, while
$\operatorname{fin}(\varphi)=0$ and $\operatorname{fin}(\mathrm{true})=1$.
Hence $\{a\}\{b\}\models\varphi$, and $\{a\}\not\models\varphi$. To
compare $\varphi$ with $\psi=\mathbf Fb$, for which
$\operatorname{prog}(\psi,\emptyset)=\psi$ and
$\operatorname{prog}(\psi,\{b\})=\mathrm{true}$, the trace $\emptyset\{b\}$
progresses the pair $(\varphi,\psi)$ to $(\mathrm{false},\mathrm{true})$,
whose end values differ, while no shorter trace distinguishes $\varphi$
from $\psi$.
\end{example}

The procedure is a breadth-first search over pairs of progressed
formulas. It starts from $(\varphi,\psi)$ with the empty trace. From a pair
reached by a trace $u$, each letter $a$ leads to the pair progressed by
$a$, reached by $ua$. The search returns $u$ as soon as the two end
values differ. It visits each pair once and does not expand a pair with
equal components, since no extension of $u$ distinguishes the formulas
there. We keep progressed formulas in a \emph{normal form}, as sets of clauses
over the subformulas of the input, so only finitely many pairs arise and
the search terminates. In the worst case it visits doubly exponentially
many pairs in $|\varphi|+|\psi|$; in practice it builds only the pairs it
reaches and stops at the first witness. The implementation bounds the
number of pairs (Section~\ref{sec:implementation}); a search that exceeds
the bound reports a failure instead of an answer.

\begin{lemma}[Correctness of \textsc{Distinguish}]
\label{lem:distinguish}
Let $\varphi$ and $\psi$ be LTLf formulas, and let \textsc{Distinguish}
be the automata-based or the progression-based procedure, without a
budget. Then $\textsc{Distinguish}(\varphi,\psi)$ terminates, and it
returns a trace if and only if some $w\in\Sigma^*$ satisfies
$(w\models\varphi)\oplus(w\models\psi)$; every returned trace satisfies
this condition. Hence it returns $\mathrm{false}$ if and only if
$\varphi$ and $\psi$ are equivalent over $\Sigma^*$. The
progression-based procedure returns a shortest such trace.
\end{lemma}
The proof is given in Appendix~\ref{app:proofs}.

\paragraph{Environments.}
For an environment given as a finite labelled transition system, the
progression-based search runs over triples of an environment state and
two progressed formulas, and follows the transitions of the environment
instead of all letters. There are finitely many triples, so the search
returns a nonempty feasible distinguishing trace if and only if one
exists. During learning, the implementation also bounds the number of
actions (Section~\ref{sec:implementation}); a search that reaches this
bound returns no answer, which preserves soundness but not completeness.

\subsection{Natural-language hints}
\label{sec:nl-hints}

A hint gives the learner a \emph{partial requirement}: a property that
every acceptable execution has, but that need not describe the whole
target. Hints never change a label. They act on
Algorithm~\ref{alg:main} in two ways only: by reordering the rivals in
\textsc{NextRival}, or by restricting the class schedule of
\textsc{VersionSpace}. An execution shown to the teacher may prompt a
hint, but the learner uses only the answer. The learner asks at two
levels (Table~\ref{tab:hint-questions}): at Level~1 the teacher selects
a question type $c$, and at Level~2 they state a requirement, which is
translated into an LTLf formula $h$.

\begin{table}[t]
\centering
\footnotesize
\caption{Questions at both levels and the formula families
$\mathcal H_c$ of the question types. Level~1 answers are Yes, No, or
Unsure; Level~2 is asked after a Yes answer. $B$ and $B'$ stand for
Boolean formulas. In our experiments, a simulated teacher answers.}
\label{tab:hint-questions}
\setlength{\tabcolsep}{2.5pt}
\begin{tabularx}{\linewidth}{
    @{}
    >{\raggedright\arraybackslash}p{0.15\linewidth}
    >{\raggedright\arraybackslash}X
    >{\raggedright\arraybackslash}X
    >{\raggedright\arraybackslash}p{0.30\linewidth}
    @{}
}
\toprule
\textbf{Type} & \textbf{Level 1 question} & \textbf{Level 2 question}
& \textbf{Patterns in $\mathcal H_c$} \\
\midrule
Undesired behaviour
& Does anything happen that should not happen?
& Which event should not happen, and when?
& $\mathbf GB$, $\mathbf G\neg B$, $\mathbf G\neg(B\land B')$ \\
\addlinespace
Missing behaviour
& Is something required missing?
& Which event is missing, and when is it required?
& $\mathbf FB$, $\mathbf G(B\rightarrow\mathbf FB')$,
  $\mathbf G(B\rightarrow\mathbf X\mathbf FB')$ \\
\addlinespace
Ordering
& Do any events happen in the wrong order?
& Which events, and in what order?
& $B\,\mathbf U\,B'$, $\mathbf F(B\land\mathbf F(B'\land\cdots))$,
  $\mathbf F(B\land\mathbf{XF}(B'\land\cdots))$,
  $\mathbf G(B\rightarrow\mathbf FB')$, $\mathbf G(B\rightarrow\mathbf{XF}B')$,
  $\mathbf G(B\rightarrow\mathbf XB')$ \\
\addlinespace
Completion
& Is any task left unfinished?
& Which task, and what would complete it?
& $\mathbf FB$, $\mathbf G(B\rightarrow\mathbf FB')$,
  $\mathbf G(B\rightarrow\mathbf{XF}B')$, $\mathbf F\mathbf GB$ \\
\addlinespace
Repetition
& Does anything happen too often or not often enough?
& Which event, and how should its repetition change?
& $\mathbf G\mathbf FB$ \\
\bottomrule
\end{tabularx}
\end{table}

\paragraph{Level 1: question types.}
Each question type $c$ stands for a set of specification
patterns~\cite{dwyer1999patterns}, and $\mathcal H_c$ is the family of
formulas that are instances of these patterns
(Table~\ref{tab:hint-questions}). For example, missing behaviour stands
for eventuality and response, with instances such as $\mathbf Fa$ and
$\mathbf G(b\rightarrow\mathbf Fa)$. The families overlap, since one
requirement can be described by different question types. The learner
uses only the question types answered with Yes.

\paragraph{Level 2: a containing language.}
The Level-2 answer names the events and conditions of a requirement. It
is translated into a formula $h$ that every acceptable execution should
satisfy, for example by a large language model; we assume the
translation is given and do not evaluate it. A correct hint satisfies
$\Target\models h$, that is, $L(\Target)\subseteq L(h)$. For example,
``every pickup must be followed by a delivery'' gives
$h=\mathbf G(\mathit{pickup}\rightarrow\mathbf F\,\mathit{deliver})$. This
hint is correct for the target of the motivating example
(Example~\ref{ex:motivating}), but it does not determine that target.

\paragraph{Using hints.}
$\textsc{AskHint}$ (line~7 of Algorithm~\ref{alg:main}) asks the Level-1
question at most once in a run. The families $\mathcal H_c$ of the
question types answered with Yes are the \emph{hinted families}. The
learner uses them in one of two ways.

\emph{Ranking.} \textsc{NextRival} sorts the rivals stably by their
distance to the hinted families: the smallest number of syntactic
changes that turn the rival into an instance of a hinted family.
Ranking changes only the order in which rivals are compared.

\emph{Pruning.} The class schedule is replaced by levels: first the
formulas of $\mathcal L_{n,d}$ in the hinted families, then those whose
root operator is the root operator of some formula in a hinted family,
and finally $\mathcal L_{n,d}$. Each level is searched by size, as in the
default schedule. A Level-2 hint $h$ adds a first level that contains
the formulas $\psi$ rejecting every trace of $T_\ell$ that $h$ rejects,
which is the check $\psi\models h$ restricted to $T_\ell$. The learner
moves to the next level only when the current level leaves the version
space empty. Since pruning changes the class schedule, its hints are
asked before the first call of \textsc{VersionSpace} (line~4) rather
than at line~7.

Hints never change a label, and within the searched class the version
space is filtered by labels only, so every hypothesis remains consistent
with $S$ whatever the accuracy of the hints
(Proposition~\ref{prop:sample-consistency}). Ranking excludes no
candidate, so an incorrect ranking hint can only lead to less useful
queries. Pruning excludes candidates: if an incorrect hint leaves a
consistent formula in a level without $\Target$, the learner ends in
that level. If every hint is correct, $\Target$ lies in the first level,
and Theorem~\ref{thm:bounded-identification} applied to that level shows
that the exhaustive rule still ends with $\Hyp\equiv\Target$.

\subsection{Certify}
\label{sec:certify}
We assess distributional error in the probably approximately correct
learning framework~\cite{valiant1984theory}, using a binomial test of a fixed
hypothesis~\cite{clopper1934use}.
The final hypothesis is frozen. $\textsc{Certify}(\Hyp,P)$ draws up to
$P$ fresh traces i.i.d.\ from a declared distribution $\mathcal D$ and
asks their labels by membership queries. It returns a flag $b$ and a
trace $w$: it fails ($b=0$) as soon as a test disagrees with $\Hyp$ and
returns that trace as the counterexample $w$; if all $P$ tests agree, it
succeeds ($b=1$) without a counterexample, and the guarantee of
Theorem~\ref{thm:pac} applies.

\section{Formal Properties}
\label{sec:properties}
The identification results concern Algorithm~\ref{alg:main} with the
exhaustive rule ($K=\infty$), the class schedule by size, no hint
pruning, and a comparison budget that is not exhausted; $\mathcal H$
denotes the largest searched class $\mathcal L_M$, and $\mathcal H/{\equiv}$ its
set of equivalence classes. Proofs are in Appendix~\ref{app:proofs}.

\begin{proposition}[Sample consistency]
\label{prop:sample-consistency}
Any hypothesis returned from a nonempty version space agrees with all labels
in $S$.
\end{proposition}

\begin{theorem}[Conditional bounded identification]
\label{thm:bounded-identification}
Let $\Target\in\mathcal H$, and suppose \textsc{Distinguish} returns a
witness for every pair of non-equivalent formulas of $\mathcal H$. Then
Algorithm~\ref{alg:main} asks at most $|\mathcal H/{\equiv}|-1$
refinement queries (line~15) and ends with $\Hyp\equiv\Target$.
\end{theorem}

Under either stopping rule, every query removes at least one equivalence
class (Section~\ref{sec:approach}), so a run asks at most
$L+|\mathcal H/{\equiv}|-1+P$ membership queries, including the initial
sample and certification; the comparison budget bounds comparisons, not
queries. The implementation bounds \textsc{Distinguish} by a number of
product states and, in environments, by a horizon
(Section~\ref{sec:implementation}); in our experiments, these bounds did
not affect any final hypothesis (Section~\ref{sec:rq-ident}).

\paragraph{Identification relative to an environment.}
In an environment $\mathcal E$, queries are restricted to its feasible
traces $L(\mathcal E)\subseteq\Sigma^+$, so formulas that differ only on
infeasible traces cannot be told apart.

\begin{definition}[Environment-relative equivalence]
Two formulas $\varphi$ and $\psi$ are equivalent
relative to an environment $\mathcal E$, written
$\varphi\equiv_{\mathcal E}\psi$, if
\(
L(\mathcal E)\cap
\big(L(\varphi)\triangle L(\psi)\big)
=\emptyset.
\)
\end{definition}

Equivalence implies $\equiv_{\mathcal E}$, but not conversely, so
$N_{\mathcal E}=|\mathcal H/{\equiv_{\mathcal E}}|\leq|\mathcal H/{\equiv}|$.
\textsc{Distinguish} then searches for a witness in
$L(\mathcal E)\cap(L(\varphi)\triangle L(\psi))$.

\begin{theorem}[Conditional environment-relative identification]
\label{thm:env-identification}
Let $\Target\in\mathcal H$, and suppose \textsc{Distinguish} returns a
feasible witness for every pair of formulas of $\mathcal H$ that are not
equivalent relative to $\mathcal E$. Then Algorithm~\ref{alg:main}
asks at most $N_{\mathcal E}-1$ refinement queries and ends with
$\Hyp\equiv_{\mathcal E}\Target$.
\end{theorem}

The environment can thus reduce the number of queries. The guarantee
needs a complete search over feasible traces: if feasible traces are
only sampled, failing to find a witness does not establish
$\equiv_{\mathcal E}$.

\paragraph{Certification.}
The following bound inverts the binomial probability of observing no
errors, as in exact binomial confidence limits~\cite{clopper1934use}.

\begin{theorem}[PAC guarantee]
\label{thm:pac}
Let $\Hyp$ be fixed before $P$ tests are drawn i.i.d.\ from $\mathcal D$.
With probability at least $1-\delta$ over the tests, some test disagrees
with $\Hyp$, or
$\operatorname{err}_{\mathcal D}(\Hyp,\Target)\le\ln(1/\delta)/P$.
\end{theorem}

The same holds for a distribution $\mathcal D_{\mathcal E}$ supported on
$L(\mathcal E)$, with
$\operatorname{err}_{\mathcal D_{\mathcal E}}(\Hyp,\Target)=
\Pr_{w\sim\mathcal D_{\mathcal E}}[(w\models\Hyp)\neq(w\models\Target)]$.
Unlike Theorem~\ref{thm:env-identification}, this bound needs no
exhaustive exploration of the environment, but it bounds the
disagreement under $\mathcal D_{\mathcal E}$, not $\equiv_{\mathcal E}$.

\section{Implementation}
\label{sec:implementation}
We implemented Algorithm~\ref{alg:main} for LTLf in Python~3.14. Each
procedure of Section~\ref{sec:approach} is a separate module, so the
enumeration-based and the SAT-based learner share every procedure except
\textsc{VersionSpace}. The code and all experiments are available at
\url{https://github.com/rajarshi008/ltl-active-learning-with-hints}.
For the pool $T_\ell$ of \textsc{Distinguish}
(Section~\ref{sec:distinguish}), we choose $\ell$ as the largest length
with at most $5{,}000$ traces in $T_\ell$. If no trace of $T_\ell$
distinguishes the two formulas, the progression-based search
runs, bounded by $10^5$ product states and, in environments, by $30$
actions; a comparison that reaches this horizon is undecided, and the
rival is skipped. For the automata-based search, we build the automata
with LTLf2DFA~1.0.2~\cite{fuggitti2019ltlf2dfa} and
MONA~1.4~\cite{elgaard1998mona}.

All experiments ran on an Apple M4 Pro with $24$\,GB of memory. 
The solver is the incremental SAT engine of Z3~5.3.3~\cite{demoura2008z3}, used through its
Python API; a run is halted after $2\cdot10^5$ solver calls or $600$\,s.
The
environments use Gymnasium~1.4.0~\cite{towers2025gymnasium} and
MiniGrid~3.2.0~\cite{chevalierboisvert2023minigrid}.

\section{Experimental Evaluation}
\label{sec:experiments}
Our evaluation studies the practical performance of the learner:
whether it identifies the target, what identification costs in queries
and time, and how natural-language hints change both. All experiments
concern LTLf over propositional traces.

\begin{description}
    \item[RQ1.] \textbf{(Identification).}
    Does the learner return a hypothesis equivalent to the target, and
    how do the stopping rule and the order of rivals affect this?

    \item[RQ2.] \textbf{(Cost).}
    How many membership queries and how much time does the learner need,
    how do they grow with the size of the target, and how much do the
    order of rivals and the implementation change them?

    \item[RQ3.] \textbf{(Natural-language hints).}
    How do correct hints affect the number of queries, candidate
    generation and the running time, and how do incorrect hints affect
    identification?
\end{description}

\subsection{Experimental setup}
\label{sec:setup}

\paragraph{Benchmarks.}
We use two benchmark sets (Table~\ref{tab:benchmarks}). The
\emph{free} set contains $20$ LTLf targets over $\AP=\{a,b,c\}$, so
$|\Sigma|=8$, with sizes $2$ to $7$ and temporal depth at most $3$; here
queries may be arbitrary traces. The \emph{environment} set contains $16$
targets in four standard benchmarks: the Office world from the
reward-machine literature~\cite{toroicarte2022reward}, the
deterministic version of Taxi~\cite{dietterich2000hierarchical} from
Gymnasium~\cite{towers2025gymnasium}, and
the DoorKey $5\times5$ and $6\times6$ layouts of
MiniGrid~\cite{chevalierboisvert2023minigrid}. Each benchmark is represented as a finite
deterministic labelled transition system with a fixed start state; states
at which the original episode ends are made absorbing. Queries are
restricted to feasible traces. Each target uses the propositions listed
in Table~\ref{tab:benchmarks}; Appendix~\ref{app:environments} describes
the environments, their adaptations and the meaning of the propositions.

\begin{table}[t]
\centering
\small
\caption{Benchmark targets. $|S|$: number of reachable environment
states. Sizes count syntax-tree nodes, with $\neg p$ counted as one node.}
\label{tab:benchmarks}
\setlength{\tabcolsep}{3pt}
\begin{tabularx}{\linewidth}{@{}l>{\raggedright\arraybackslash}X@{}}
\toprule
\multicolumn{2}{@{}l}{\textbf{Free traces}, $\AP=\{a,b,c\}$} \\
\midrule
Size 2 & $\mathbf Fa$, $\mathbf Ga$ \\
Size 3 & $\mathbf{FG}a$, $\mathbf{GF}a$, $a\,\mathbf U\,b$ \\
Size 5 & $\mathbf Fa\rightarrow\mathbf Fb$, $\mathbf G(a\rightarrow\mathbf Fb)$,
  $\mathbf G(a\rightarrow\mathbf Xb)$, $\mathbf F(a\land\mathbf Fb)$,
  $a\,\mathbf U\,(b\,\mathbf U\,c)$, $\mathbf Fa\land\mathbf G\neg b$,
  $\mathbf F(a\land\mathbf Xb)$ \\
Size 6 & $\mathbf G(a\rightarrow b\,\mathbf U\,c)$,
  $a\rightarrow\mathbf X(b\,\mathbf U\,c)$,
  $\neg a\,\mathbf U\,(b\land\mathbf Fc)$,
  $\mathbf G(\neg a\lor\mathbf{XG}\neg b)$ \\
Size 7 & $\mathbf G((a\land b)\rightarrow\mathbf Xc)$,
  $\mathbf{GF}a\land\mathbf{GF}b$,
  $\mathbf G(a\rightarrow\mathbf X(a\,\mathbf U\,b))$,
  $\mathbf F(a\land\mathbf F(b\land c))$ \\
\midrule
\multicolumn{2}{@{}l}{\textbf{Environments}} \\
\midrule
Office, $|S|=108$ &
  $\{f,g,n\}$: $\mathbf G\neg n$, $\neg n\,\mathbf U\,f$,
  $\mathbf F(f\land\mathbf Fg)$, $\mathbf Ff\land\mathbf G\neg n$,
  $\neg n\,\mathbf U\,(f\land\mathbf Fg)$;
  $\{e,g,n\}$: $\mathbf F(e\land\mathbf Fg)$;
  $\{a,b,c\}$: $\mathbf F(a\land\mathbf Fb)$ \\
Taxi, $|S|=101$ & $\{p,g\}$: $\mathbf Fg$, $\mathbf F(p\land\mathbf Fg)$,
  $\mathbf G(p\rightarrow\mathbf Fg)$ \\
DoorKey $5{\times}5$, $|S|=62$ & $\{k,o,g\}$: $\mathbf Fg$,
  $\neg o\,\mathbf U\,k$, $\mathbf F(o\land\mathbf Fg)$,
  $\mathbf G(k\rightarrow\mathbf Fo)$ \\
DoorKey $6{\times}6$, $|S|=154$ & $\{k,o,g\}$: $\mathbf Fg$,
  $\mathbf F(k\land\mathbf Fg)$ \\
\bottomrule
\end{tabularx}
\end{table}

\paragraph{Learner variants.}
Every run uses the enumeration-based version space with the class
schedule by size (Section~\ref{sec:version-space}) and the
progression-based \textsc{Distinguish} after the short-trace pool
$T_\ell$ (Section~\ref{sec:distinguish}). For a target $\Target$ we
search $\mathcal L_{n,d}$ with $n=|\Target|+1$ and $d=2$, or $d=3$ for
the two free targets of temporal depth $3$. The bound $n$ is thus assumed
to be known up to one, a limitation we return to in
Section~\ref{sec:conclusion}.
$\textsc{Initialize}$ labels $L=10$ traces: uniformly random traces of
length $1$ to $6$ for free targets, and rollouts with uniformly random
actions and $1$ to $10$ labels for environments. The first hypothesis is
a smallest consistent candidate; when a witness refutes the hypothesis,
the compared rival replaces it. The comparison budget is $100$ (under the
exhaustive rule, only comparisons that lead to a query count) and was
never exhausted.

A variant of Algorithm~\ref{alg:main} is fixed by its stopping rule and
its order of rivals (Section~\ref{sec:next-rival}):
\begin{itemize}
    \item \emph{Stopping rule.} \emph{Top-$3$} is the top-$K$ rule with
    $K=3$: a round compares $\Hyp$ with the three highest-ranked rivals,
    and refinement ends after a round without a query.
    \emph{Exhaustive} is $K=\infty$: a round compares $\Hyp$ with the
    rivals in order, removes those equivalent to $\Hyp$ (relative to
    $\mathcal E$ in environments) without a query, and refinement ends
    only when no remaining candidate is distinguishable from $\Hyp$. Only
    the exhaustive rule is covered by the correctness argument of
    Section~\ref{sec:approach}.
    \item \emph{Order of rivals.} \emph{Enumeration} (smallest first);
    \emph{semantic}, closest first under $d_{\rm sem}$ on $r=200$ probe
    traces ($200$ rollouts in environments), and its reverse
    \emph{semantic-far}; and \emph{split}.
\end{itemize}
RQ1 and RQ2 report these four orders under both rules without hints
($1{,}440$ runs); four further orders ($1{,}440$ runs) are omitted for
space. RQ3 runs enumeration order under both rules with
eight hint settings ($2{,}660$ runs), one of them repeating the no-hint
configuration of RQ1. For
RQ2, we also compare two alternative implementations on the free
targets under the exhaustive rule: the automata-based
\textsc{Distinguish} with MONA, and the SAT-based version space.
Each configuration runs with seeds $41$ to $45$ ($100$ free and $80$
environment runs); the certification results come from a rerun of the
$720$ top-$3$ runs that reproduces Table~\ref{tab:rq-ident} exactly.
Section~\ref{sec:implementation} gives the hardware.

\paragraph{Teacher and metrics.}
A simulated teacher answers membership queries by evaluating $\Target$.
A run is \emph{successful} (Succ.) if the final hypothesis satisfies
$\Hyp\equiv\Target$ (free targets) or $\Hyp\equiv_{\mathcal E}\Target$
(environments); both are decided exactly by a complete product search
that is used only for evaluation. A run is \emph{exact} if
$\Hyp=\Target$ up to notation. MQ counts all labelled traces, including
the $L=10$ initial ones. Eq.\ counts the comparisons under the exhaustive
rule that find a rival equivalent to $\Hyp$; they remove the rival
without a query. Cand.\ counts the formulas generated by the enumerator,
and Time is the wall-clock time of a run.

\subsection{RQ1: Identification}
\label{sec:rq-ident}

\begin{table*}[t]
\centering
\small
\caption{RQ1 and RQ2: identification and cost by order of rivals and
stopping rule. All runs use the enumeration-based version space, the
progression-based \textsc{Distinguish} and no hints. Top-$3$: top-$K$
rule with $K=3$; Exhaustive: $K=\infty$. Succ.: runs with
$\Hyp\equiv\Target$ (free, out of $100$) or
$\Hyp\equiv_{\mathcal E}\Target$ (environments, out of $80$). Exact:
runs with $\Hyp=\Target$ up to notation; under the exhaustive rule, $85$
free and $30$ environment runs for every order. MQ: mean membership
queries per run, including the $10$ initial traces; top-$3$ means
include failed runs. Eq.: mean comparisons per run under the exhaustive
rule that found a rival equivalent to $\Hyp$.}
\label{tab:rq-ident}
\setlength{\tabcolsep}{4pt}
\begin{tabular}{@{}lrrrrrrrrrrrr@{}}
\toprule
& \multicolumn{6}{c}{\textbf{Free traces}}
& \multicolumn{6}{c}{\textbf{Environments}} \\
\cmidrule(lr){2-7}\cmidrule(l){8-13}
& \multicolumn{3}{c}{Top-$3$} & \multicolumn{3}{c}{Exhaustive}
& \multicolumn{3}{c}{Top-$3$} & \multicolumn{3}{c}{Exhaustive} \\
\cmidrule(lr){2-4}\cmidrule(lr){5-7}\cmidrule(lr){8-10}\cmidrule(l){11-13}
\textbf{Order of rivals} & Succ. & Exact & MQ & Succ. & MQ & Eq.
& Succ. & Exact & MQ & Succ. & MQ & Eq. \\
\midrule
Enumeration         &  92 & 77 & 15.5 & 100 & 16.2 & 14.2 & 40 & 20 &  8.7 & 80 & 12.5 & 310 \\
Semantic (closest)  &  64 & 56 & 13.5 & 100 & 15.9 & 20.1 & 40 & 20 &  8.6 & 80 & 12.2 & 310 \\
Semantic (farthest) & 100 & 85 & 16.1 & 100 & 16.1 & 11.9 & 41 & 20 &  9.5 & 80 & 12.5 & 308 \\
Split               & 100 & 85 & 15.7 & 100 & 15.7 & 11.8 & 64 & 25 & 10.6 & 80 & 12.4 & 296 \\
\bottomrule
\end{tabular}
\end{table*}

\paragraph{Exhaustive rule.}
Table~\ref{tab:rq-ident} shows that under the exhaustive rule every run
succeeds, for every order of rivals: $100$ of $100$ free and $80$ of
$80$ environment runs. This is the outcome predicted by the correctness
argument of Section~\ref{sec:approach}, since every target lies in the
searched class $\mathcal L_{n,d}$. In environments, the $30$-action
horizon of \textsc{Distinguish} (Section~\ref{sec:implementation}) could
leave a distinguishable rival in the version space; it did not affect
any final hypothesis. The remaining hypotheses that are not the target
formula itself (Exact in Table~\ref{tab:rq-ident}) are equivalent to it
on the relevant trace domain; in Taxi, for example, $\mathbf Fg$ and
$\mathbf F(p\land\mathbf Fg)$ agree on all feasible traces because
delivery requires a prior pickup.

\paragraph{Top-$3$ rule.}
Under the top-$3$ rule, identification depends on which rivals are compared first. On free traces, semantic ordering succeeded in $64$ of $100$ runs, while semantic-far and split succeeded in all $100$. All free-trace failures involved targets of size $5$ to $7$. In environments, success ranged from $40$ to $64$ of $80$ runs.

Every failed run stopped because none of the three shortlisted rivals yielded a query, even though distinguishable candidates remained. No run exhausted the comparison budget. Closest-first ordering often shortlists equivalent rewritings, which have $d_{\rm sem}=0$ on the probe set. Semantic-far and split avoided premature stopping on free traces. Equivalent rivals are more common in environments. In Office world, for example, a formula that requires the agent to start in a labelled cell is equivalent to $\mathrm{false}$ on feasible traces. With enumeration order, only $5$ of $35$ Office world runs succeeded.

\paragraph{Certification.}
Certification detected all identification failures on free traces, but few in environments. We tested each top-$3$ hypothesis using \textsc{Certify} (Section~\ref{sec:certify}). Free traces had geometric lengths with mean $5$ and uniform letters. Environment rollouts had geometric step counts with mean $10$ and uniform actions. Table~\ref{tab:rq-ident} excludes certification queries.

On free traces, certification detected $38$ of $44$ failures within $30$ tests and all $44$ within $60$ tests ($\epsilon=\delta=0.05$). In environments, it detected $3$ of $135$ failures within $60$ tests and $21$ within $300$ ($15/15$ in DoorKey, $6/105$ in Office world, and $0/15$ in Taxi). All equivalent hypotheses passed. For undetected hypotheses, the estimated error under $\mathcal D_{\mathcal E}$ was at most $0.2\%$ over $20{,}000$ rollouts. Thus, low error under the test distribution does not imply equivalence on feasible traces.

A failed test can be added to $S$ to resume refinement, approximating an equivalence query by random sampling~\cite{angluin1988queries}. This strategy offered limited benefit here. On free traces, exhaustive refinement identified every target with at most $2.4$ more queries on average than the top-$3$ rule. In environments, random tests rarely found a counterexample.

\smallskip\noindent\emph{Answer to RQ1.}
The exhaustive rule identified every benchmark target up to $\equiv$ on free traces or $\equiv_{\mathcal E}$ in environments, regardless of rival order. With the top-$3$ rule, success ranged from $64$ to $100$ of $100$ free runs and from $40$ to $64$ of $80$ environment runs. Closest-first performed worst on free traces because it often shortlisted rivals equivalent to the hypothesis. In environments, split clearly outperformed the other orders. Certification detected all failures on free traces but few in environments, where random rollouts rarely exposed disagreements.

\subsection{RQ2: Cost}
\label{sec:rq-cost}

\begin{table}[t]
\centering
\small
\caption{RQ2: cost by target size under the exhaustive rule with
enumeration order (no hints). $N$: runs; MQ: mean (minimum--maximum); Time: mean
seconds per run; Max: maximum seconds per run; Cand.: mean number of
generated formulas, in thousands.}
\label{tab:rq-cost-size}
\setlength{\tabcolsep}{2pt}
\begin{tabular}{@{}lrlrrrrlr@{}}
\toprule
& \multicolumn{5}{c}{\textbf{Free traces}} & \multicolumn{3}{c}{\textbf{Environments}} \\
\cmidrule(lr){2-6}\cmidrule(l){7-9}
$|\Target|$ & $N$ & MQ & Time & Max & Cand. & $N$ & MQ & Time \\
\midrule
2 & 10 & 10.6 (10--13) & 0.01 &  0.03 &     0.2 & 20 &  8.7 (5--11)  & 0.02 \\
3 & 15 & 11.4 (10--13) & 0.01 &  0.02 &     1.0 & 10 & 10.5 (7--13)  & 0.02 \\
5 & 35 & 15.7 (10--20) & 0.10 &  0.11 &    55.8 & 45 & 14.0 (9--21)  & 0.19 \\
6 & 20 & 18.6 (14--22) & 1.07 &  1.32 &     536 &  5 & 18.4 (17--19) & 1.02 \\
7 & 20 & 20.9 (13--32) & 7.94 & 10.70 & 3{,}944 & -- & --            & --   \\
\bottomrule
\end{tabular}
\end{table}

\begin{table}[t]
\centering
\footnotesize
\caption{RQ2: three implementations on the free targets (exhaustive
rule, enumeration order, no hints). Progr.: enumerated version space
and progression-based \textsc{Distinguish}, as in all other runs; MONA:
the same with automata-based \textsc{Distinguish}; SAT: SAT-based
version space with progression-based \textsc{Distinguish}. Time: mean
seconds per run; Mem.: mean peak memory in MB. Each run has its own
process, which adds about $0.7$\,s, so times are not comparable with
Table~\ref{tab:rq-cost-size}. Rows for $n=9,10$: the four size-$7$
targets with seeds $41$ and $42$ and a limit of $600$\,s (one run each
for Progr.\ and MONA); the SAT entry gives the runs that finished.}
\label{tab:rq-impl}
\setlength{\tabcolsep}{2.5pt}
\begin{tabular}{@{}rrrrrrrr@{}}
\toprule
& & \multicolumn{2}{c}{\textbf{Progr.}} & \multicolumn{2}{c}{\textbf{MONA}}
& \multicolumn{2}{c}{\textbf{SAT}} \\
\cmidrule(lr){3-4}\cmidrule(lr){5-6}\cmidrule(l){7-8}
$|\Target|$ & $n$ & Time & Mem. & Time & Mem. & Time & Mem. \\
\midrule
2 & 3 & 0.83 & 281 & 0.78 & 281 & 0.89 & 315 \\
3 & 4 & 0.79 & 281 & 0.80 & 281 & 0.98 & 316 \\
5 & 6 & 0.87 & 290 & 0.97 & 290 & 1.63 & 356 \\
6 & 7 & 1.81 & 350 & 1.98 & 352 & 3.03 & 415 \\
7 & 8 & 8.82 & 702 & 9.92 & 728 & 22.35 & 486 \\
7 & 9 & 70.2 & 3{,}471 & 69.6 & 3{,}858 & 2/8 & 579 \\
7 & 10 & -- & -- & -- & -- & 0/8 & 709 \\
\midrule
\multicolumn{2}{@{}l}{Exact} & \multicolumn{2}{c}{85} & \multicolumn{2}{c}{85} & \multicolumn{2}{c}{50} \\
\multicolumn{2}{@{}l}{MQ} & \multicolumn{2}{c}{16.2} & \multicolumn{2}{c}{16.2} & \multicolumn{2}{c}{16.1} \\
\bottomrule
\end{tabular}
\end{table}

\paragraph{Queries.}
With the exhaustive rule and enumeration order, a run asks $16.2$
queries on average on free targets (standard deviation $4.5$, range $10$
to $32$) and $12.5$ in environments (standard deviation $3.8$, range $5$
to $21$), that is, $6.2$ and $2.5$ queries beyond the initial sample.
Table~\ref{tab:rq-cost-size} shows that the number of queries grows with
the size of the target, from $10.6$ for size $2$ to $20.9$ for size $7$
on free targets.

\paragraph{Order of rivals and stopping rule.}
Under the exhaustive rule, the order of rivals changes the mean number
of queries by at most $3\%$ on free targets ($15.7$ for split to
$16.2$ for enumeration order) and $2\%$ in environments ($12.2$ to
$12.5$; Table~\ref{tab:rq-ident}). Compared with the top-$3$ rule under
the same order, the exhaustive rule asks $0$ to $2.5$ more queries on
free targets and $1.8$ to $3.8$ more in environments. These are the
queries that the top-$3$ rule omits; for instance, the low mean of
semantic ordering under the top-$3$ rule ($13.5$) reflects runs that end
early, $36$ of them without identifying the target. Equivalence checks
cost no queries: the exhaustive rule removes $11.8$ to $20.1$ equivalent
rivals per run on free targets, most under closest-first orders, and
$296$ to $310$ in environments.

\paragraph{Running time.}
The mean running time per run hardly depends on the variant: $1.83$ to
$1.84$\,s on free targets and $0.15$ to $0.18$\,s in environments, for
every order and both rules. It grows with the size bound $n$, together
with the number of generated formulas: for size-$7$ targets ($n=8$), the
enumerator generates up to $5.1$ million formulas and a run takes
$7.9$\,s on average and at most $10.7$\,s. We did not include targets of
size $8$. With $n=9$, a single run (target
$\mathbf G((a\land b)\rightarrow\mathbf Xc)$, seed $41$) generated
$31.4$ million formulas and took $70$\,s and $3.4$\,GB of memory
(Table~\ref{tab:rq-impl}).

\begin{table*}[t]
\centering
\small
\caption{RQ3: effect of hints. All runs use enumeration order of
rivals and the enumeration-based version space. Succ.: successful runs
under the top-$3$ and the exhaustive rule. MQ, Cand.\ (generated
formulas, in thousands) and Time (seconds) are means under the
exhaustive rule. Rows marked $^\dagger$ cover only the targets that
receive a Level-2 hint ($70$ free and $55$ environment runs); None$^\dagger$
is the run without hints on these targets.}
\label{tab:rq-hints}
\setlength{\tabcolsep}{4pt}
\begin{tabular}{@{}lrrrrrrrrrr@{}}
\toprule
& \multicolumn{5}{c}{\textbf{Free traces}} & \multicolumn{5}{c}{\textbf{Environments}} \\
\cmidrule(lr){2-6}\cmidrule(l){7-11}
& \multicolumn{2}{c}{Succ.} & \multicolumn{3}{c}{Exhaustive}
& \multicolumn{2}{c}{Succ.} & \multicolumn{3}{c}{Exhaustive} \\
\cmidrule(lr){2-3}\cmidrule(lr){4-6}\cmidrule(lr){7-8}\cmidrule(l){9-11}
\textbf{Hints} & Top-$3$ & Exh. & MQ & Cand. & Time & Top-$3$ & Exh. & MQ & Cand. & Time \\
\midrule
None                         & 92/100 & 100/100 & 16.2 &    916 & 1.83 & 40/80 & 80/80 & 12.5 & 57 & 0.18 \\
Level 1, ranking             & 95/100 & 100/100 & 16.0 &    916 & 1.84 & 64/80 & 80/80 & 12.4 & 57 & 0.20 \\
Level 1, pruning             & 94/100 & 100/100 & 14.6 &    710 & 1.56 & 54/80 & 80/80 & 10.5 & 46 & 0.12 \\
\midrule
None$^\dagger$               & 66/70 & 70/70 & 16.0 &    964 & 1.94 & 25/55 & 55/55 & 13.7 & 77 & 0.23 \\
Level 2$^\dagger$            & 67/70 & 70/70 & 14.1 &    964 & 1.96 & 25/55 & 55/55 & 11.9 & 77 & 0.23 \\
Level 1 pruning + Level 2$^\dagger$ & 67/70 & 70/70 & 14.0 & 964 & 1.96 & 45/55 & 55/55 & 11.9 & 77 & 0.27 \\
\midrule
Incorrect Level 1, ranking   & 93/100 & 100/100 & 15.7 &    916 & 1.84 & 45/80 & 80/80 & 12.6 & 57 & 0.20 \\
Incorrect Level 1, pruning   & 67/100 &  74/100 & 13.7 &    369 & 0.76 & 43/80 & 43/80 &  8.8 & 11 & 0.04 \\
Incorrect Level 2            & 53/100 &  58/100 & 15.1 & 1{,}290 & 2.63 & -- & -- & -- & -- & -- \\
\bottomrule
\end{tabular}
\end{table*}

\paragraph{Alternative implementations.}
Table~\ref{tab:rq-impl} compares the default implementation with the
automata-based \textsc{Distinguish} and the SAT-based version space. All
three learn every free target, with $16.1$ to $16.2$ queries on average.
With MONA, the hypotheses and queries are identical in all $100$ runs,
and runs take at most $13\%$ longer: most comparisons end in the
short-trace pool, and only the others build automata. The SAT-based
version space is slower at every size, $22.4$\,s against $8.8$\,s for
size-$7$ targets. With $n=9$, only $2$ of $8$ runs finish within
$600$\,s, and with $n=10$ none does, although the hypothesis is already
equivalent to the target in all $16$ runs: the remaining time is spent
blocking encodings, since a formula has one encoding per order of its
nodes and the exhaustive rule must exclude all of them. In return, the
SAT-based construction stores no formulas and needs $0.6$\,GB with
$n=9$, against $3.4$\,GB. It also returns the target formula itself
less often ($50$ against $85$ of $100$ runs), since it does not apply
the enumerator's simplifications.

\smallskip\noindent\emph{Answer to RQ2.}
The exhaustive rule needs $16.2$ queries per free and $12.5$ per
environment target on average, growing with target size and barely
affected by the order of rivals. Running time grows with the size bound,
reaching about a minute and $3.4$\,GB for $n=9$; neither alternative
implementation lowers this cost.

\subsection{RQ3: Natural-language hints}
\label{sec:rq-hints}

\paragraph{Simulated hints.}
At Level~1, the simulated teacher selects the question types $c$ for which $\Target\in\mathcal H_c$. It finds a match for $10$ of $20$ free targets and $14$ of $16$ environment targets, and answers \emph{unsure} otherwise. At Level~2, it selects the strongest template instance $h$ that is entailed by $\Target$ but is neither valid nor equivalent to $\Target$. Entailment is checked exactly. For example, $\mathbf G(a\rightarrow\mathbf Xb)$ receives the hint $\mathbf G(a\rightarrow\mathbf Fb)$. Such hints exist for $14$ free and $11$ environment targets.

For each target, we also generate an incorrect hint at each level. An incorrect Level-1 hint selects a type $c$ with $\Target\notin\mathcal H_c$. An incorrect Level-2 hint selects a template instance not entailed by $\Target$, minimising the number of traces in $T_\ell$ that satisfy $\Target$ but violate $h$. We use Level-1 hints for ranking or pruning and Level-2 hints for pruning (Section~\ref{sec:nl-hints}), with enumeration as the base rival order and $\ell=4$ for free targets. We report incorrect Level-2 results only for free traces. In environments, these hints either hold on all feasible traces accepted by the target or have counterexamples only on traces of $14$ to $26$ labels.

\paragraph{Correct hints.}
Ranking with Level-1 hints barely changes the mean number of membership queries, but raises top-$3$ success in environments from $40$ to $64$ of $80$ runs by bringing distinguishable rivals into the shortlist (Table~\ref{tab:rq-hints}). Pruning with Level-1 hints reduces membership queries by $10\%$ on free traces ($16.2$ to $14.6$) and $16\%$ in environments ($12.5$ to $10.5$). It also reduces candidate generation by $22\%$ and $20\%$, respectively, and free-trace running time by $15\%$.

On targets that receive a Level-2 hint, pruning reduces mean membership queries from $16.0$ to $14.1$ on free traces and from $13.7$ to $11.9$ in environments. It does not reduce candidate generation because the hint is checked after enumeration. Combining both levels yields little further query savings, but raises top-$3$ success in environments from $25$ to $45$ of $55$ runs relative to the no-hint baseline. The exhaustive rule succeeds in every run with correct hints.

\paragraph{Incorrect hints.}
Incorrect ranking hints leave exhaustive identification unchanged ($100$ of $100$ free runs and $80$ of $80$ environment runs) and change the mean number of membership queries by at most $3\%$. Incorrect pruning hints can prevent identification even under the exhaustive rule. With Level-1 pruning, success drops to $74$ of $100$ free runs and $43$ of $80$ environment runs. With Level-2 pruning, it drops to $58$ of $100$ free runs.

The search widens only if the current level has no consistent candidate. An incorrect hint can therefore keep the search in a level that contains a sample-consistent formula but no formula equivalent to the target. For example, with $\Target=\mathbf{GF}a$ and the incorrect hint $h=\mathbf Fa$, the learner returned $\mathbf{FG}a$. The returned formula differs from $\Target$ only on the empty trace.

\smallskip\noindent\emph{Answer to RQ3.}
Correct pruning hints reduce queries by up to $16\%$ and candidate generation by about a fifth; ranking improves top-$3$ identification in environments. Incorrect ranking hints are harmless, but incorrect pruning hints lower exhaustive success to $54$--$74\%$.

\section{Conclusion}
\label{sec:conclusion}
We presented an active-learning framework for LTLf in which the teacher
only labels traces and answers natural-language questions. The learner
identified every benchmark target with few queries, and hints reduced
the queries further. Future work includes a study with human users and
noisy labels, translating hints with large language models, removing
the assumed bound on the target size, and scaling to larger formulas.

\section*{Acknowledgments}
This work is partially funded by the European Union (under grant agreement ID 101212818). Views and opinions expressed are however those of the author(s) only and do not necessarily reflect those of the European Union or European Health and Digital Executive Agency (HADEA). Neither the European Union nor the granting authority can be held responsible for them.

\bibliographystyle{plainnat}
\bibliography{references}

\clearpage
\appendix
\section{Distinguishing by formula progression}
\label{app:progression}
This appendix defines the progression-based \textsc{Distinguish} of
Section~\ref{sec:distinguish-prog} and proves it correct.

\paragraph{Negation normal form.}
As usual for LTLf~\cite{degiacomo2013linear}, we use formulas in negation
normal form (NNF), with the weak next $\widetilde{\mathbf X}$
($w,i\models\widetilde{\mathbf X}\varphi$ iff $i+1\geq|w|$ or
$w,i+1\models\varphi$) and release
$\varphi\,\mathbf R\,\psi=\neg(\neg\varphi\,\mathbf U\,\neg\psi)$ as the
duals of $\mathbf X$ and $\mathbf U$. Under the endpoint semantics of the
preliminaries, $\varepsilon\models\neg p$ and $\varepsilon\not\models p$.
A formula is \emph{basic} if it is a literal or its root operator is
$\mathbf X$, $\widetilde{\mathbf X}$, $\mathbf U$ or $\mathbf R$. We write
$\mathbf F\mathrm{true}$ for $\mathrm{true}\,\mathbf U\,\mathrm{true}$, which holds exactly on
nonempty words, and $\mathbf G\mathrm{false}$ for $\mathrm{false}\,\mathbf R\,\mathrm{false}$, which
holds exactly on $\varepsilon$.

\begin{definition}[Progression]
\label{def:progression-app}
For an NNF formula $\varphi$ and a letter $a\in\Sigma$, the formula
$\operatorname{prog}(\varphi,a)$ is defined by
$\operatorname{prog}(\mathrm{true},a)=\mathrm{true}$, $\operatorname{prog}(\mathrm{false},a)=\mathrm{false}$ and
\begin{align*}
\operatorname{prog}(p,a)&=\mathrm{true}\text{ if }p\in a,\text{ and }\mathrm{false}\text{ otherwise},\\
\operatorname{prog}(\neg p,a)&=\mathrm{false}\text{ if }p\in a,\text{ and }\mathrm{true}\text{ otherwise},\\
\operatorname{prog}(\varphi\land\psi,a)&=
  \operatorname{prog}(\varphi,a)\land\operatorname{prog}(\psi,a),\\
\operatorname{prog}(\varphi\lor\psi,a)&=
  \operatorname{prog}(\varphi,a)\lor\operatorname{prog}(\psi,a),\\
\operatorname{prog}(\mathbf X\varphi,a)&=\varphi\land\mathbf F\mathrm{true},\\
\operatorname{prog}(\widetilde{\mathbf X}\varphi,a)&=\varphi\lor\mathbf G\mathrm{false},\\
\operatorname{prog}(\varphi\,\mathbf U\,\psi,a)&=
  \operatorname{prog}(\psi,a)\\
  &\quad\lor\big(\operatorname{prog}(\varphi,a)\land\varphi\,\mathbf U\,\psi\big),\\
\operatorname{prog}(\varphi\,\mathbf R\,\psi,a)&=
  \operatorname{prog}(\psi,a)\\
  &\quad\land\big(\operatorname{prog}(\varphi,a)\lor\varphi\,\mathbf R\,\psi\big).
\end{align*}
For words, $\operatorname{prog}(\varphi,\varepsilon)=\varphi$ and
$\operatorname{prog}(\varphi,ua)=
\operatorname{prog}(\operatorname{prog}(\varphi,u),a)$.
The \emph{end value} $\operatorname{fin}(\varphi)\in\{0,1\}$ is $1$ for
$\mathrm{true}$, $\neg p$, $\widetilde{\mathbf X}\varphi$ and
$\varphi\,\mathbf R\,\psi$; it is $0$ for $\mathrm{false}$, $p$, $\mathbf X\varphi$
and $\varphi\,\mathbf U\,\psi$; and it is the minimum (maximum) of the
operands' values for $\land$ ($\lor$).
\end{definition}

\begin{lemma}[Progression]
\label{lem:progression-app}
For every NNF formula $\varphi$, letter $a\in\Sigma$ and word
$w\in\Sigma^*$:
(i)~$\varepsilon\models\varphi$ iff $\operatorname{fin}(\varphi)=1$; and
(ii)~$aw\models\varphi$ iff $w\models\operatorname{prog}(\varphi,a)$.
Consequently, $uw\models\varphi$ iff
$w\models\operatorname{prog}(\varphi,u)$ for all $u,w\in\Sigma^*$.
\end{lemma}
\begin{proof}
By induction on $\varphi$, as for progression over infinite
traces~\cite{bacchus2000using}. Part~(i) evaluates the semantic clauses
at the end of the word. The only case of~(ii) specific to finite words is
$\mathbf X$ (dually $\widetilde{\mathbf X}$): $aw\models\mathbf X\varphi$
iff $w\neq\varepsilon$ and $w\models\varphi$, that is,
$w\models\varphi\land\mathbf F\mathrm{true}$. The statement for words follows by
induction on $|u|$.
\end{proof}

\paragraph{Normal form.}
To keep progressed formulas from growing, the implementation stores each
one in disjunctive normal form over basic formulas: a set of clauses, each
a set of basic formulas read as a conjunction ($\mathrm{true}$ is the set
containing the empty clause, $\mathrm{false}$ the empty set). The normalisation
$\mathcal N$ (a)~removes clauses that contain $p$ and $\neg p$,
(b)~removes clauses that contain another clause, and (c)~replaces a
disjunction with unit clauses $\{p\}$ and $\{\neg p\}$ by $\mathrm{true}$. Rules (a)
and (c) are valid over $\Sigma^*$ because exactly one of $p$ and $\neg p$
holds at every position, including the end of the word, so
$\mathcal N(\chi)\equiv\chi$. We write
$\operatorname{prog}_{\mathcal N}(\chi,a)=\mathcal N(\operatorname{prog}(\chi,a))$
and extend it to words as above.

\begin{lemma}[Finitely many progressed formulas]
\label{lem:prog-finite-app}
Let $B(\varphi)$ be the set of basic subformulas of the NNF of $\varphi$,
together with $\mathbf F\mathrm{true}$ and $\mathbf G\mathrm{false}$. For every
$u\in\Sigma^*$, the formula
$\operatorname{prog}_{\mathcal N}(\mathcal N(\varphi),u)$ is a set of
subsets of $B(\varphi)$ none of which contains another. Hence at most
$2^{2^{|B(\varphi)|}}$ distinct formulas are reached from $\varphi$ by
progression, and $|B(\varphi)|\leq2|\varphi|+2$.
\end{lemma}
\begin{proof}
By Definition~\ref{def:progression-app}, progression only reintroduces
the formula itself ($\mathbf U$, $\mathbf R$), its operand ($\mathbf X$,
$\widetilde{\mathbf X}$), $\mathbf F\mathrm{true}$ or $\mathbf G\mathrm{false}$, so by induction
on $\varphi$ and $|u|$ every progressed formula is a combination of
elements of $B(\varphi)\cup\{\mathrm{true},\mathrm{false}\}$ by $\land$ and $\lor$.
$\mathcal N$ turns it into a set of clauses over $B(\varphi)$ in which,
by rule~(b), no clause contains another. Each subformula of $\varphi$
yields at most two subformulas of the NNF, one per polarity.
\end{proof}
The bound counts sets of clauses, not languages: equivalent formulas
with different normal forms only enlarge the search.

\begin{algorithm}[tb]
\caption{Progression-based distinguishing procedure
$\textsc{DistinguishProg}$}
\label{alg:distinguish-prog-app}
\begin{algorithmic}[1]
\Require LTLf formulas $\varphi$ and $\psi$; a fixed order on $\Sigma$
\Ensure A shortest word $w\in\Sigma^*$ with
  $(w\models\varphi)\oplus(w\models\psi)$, or $\mathrm{false}$ if none exists
\State $s\gets(\mathcal N(\varphi),\mathcal N(\psi))$
\If{$s_1=s_2$}
  \State \Return $\mathrm{false}$
\EndIf
\State $Q\gets\langle(s,\varepsilon)\rangle$; $V\gets\{s\}$
  \Comment{FIFO queue, visited pairs}
\While{$Q\neq\langle\rangle$}
  \State $((s_1,s_2),u)\gets\textsc{Dequeue}(Q)$
  \If{$\operatorname{fin}(s_1)\neq\operatorname{fin}(s_2)$}
    \State \Return $u$
  \EndIf
  \ForAll{$a\in\Sigma$ in the fixed order}
    \State $t\gets(\operatorname{prog}_{\mathcal N}(s_1,a),
      \operatorname{prog}_{\mathcal N}(s_2,a))$
    \If{$t_1\neq t_2$ and $t\notin V$}
      \State $V\gets V\cup\{t\}$; $\textsc{Enqueue}(Q,(t,ua))$
    \EndIf
  \EndFor
\EndWhile
\State \Return $\mathrm{false}$
\end{algorithmic}
\end{algorithm}

\begin{lemma}[Correctness of \textsc{DistinguishProg}]
\label{lem:distinguish-prog-app}
For all LTLf formulas $\varphi$ and $\psi$,
Algorithm~\ref{alg:distinguish-prog-app} terminates after at most
$2^{2^{|B(\varphi)|}}\cdot2^{2^{|B(\psi)|}}$ iterations of its main loop.
It returns a word if and only if some $w\in\Sigma^*$ satisfies
$(w\models\varphi)\oplus(w\models\psi)$, and every returned word is the
first such word in length-lexicographic order. In particular, it returns
$\mathrm{false}$ if and only if $\varphi$ and $\psi$ are equivalent over
$\Sigma^*$.
\end{lemma}
\begin{proof}
By Lemma~\ref{lem:progression-app} and $\mathcal N(\chi)\equiv\chi$, a
pair $(s_1,s_2)$ enqueued with word $u$ satisfies, for every $v$,
$uv\models\varphi$ iff $v\models s_1$ and $uv\models\psi$ iff
$v\models s_2$. With $v=\varepsilon$ and part~(i), every returned word
distinguishes the formulas; and a pair with equal components has no
distinguishing extension, so neither pruning it nor the check in line~2
loses a witness. Each pair is enqueued at most once, so termination
follows from Lemma~\ref{lem:prog-finite-app}. Since the search is
breadth-first and extends words in the fixed letter order, the first
distinguishing word it reaches is the first in length-lexicographic
order.
\end{proof}

\paragraph{Trace domain and environments.}
The search may return $\varepsilon$, a valid query under
$\mathcal W=\Sigma^*$. In an environment given as a finite labelled
transition system, it runs instead over triples of an environment state
and two progressed formulas, expands them along the transitions of the
environment, and returns nonempty feasible traces. By
Lemma~\ref{lem:prog-finite-app} there are finitely many triples, and the
argument above shows that the search returns a feasible distinguishing
trace if and only if one exists. During learning, the number of actions
is bounded (Section~\ref{sec:implementation}); a search that reaches the
bound returns no answer, which preserves soundness but not completeness.

\section{Proofs}
\label{app:proofs}
This appendix proves the results of the main text whose proofs are not
given there; each proof refers to the statement it proves.

\begin{proof}[Proof of Lemma~\ref{lem:distinguish}]
\emph{Automata.} Both translations and the emptiness test terminate.
By correctness of the LTLf-to-automaton
translation~\cite{degiacomo2013linear}, with the stated empty-word
convention, and the Boolean product
construction~\cite{rabin1959finite},
$L(\mathcal A_\Delta)=\{w\in\Sigma^*\mid
(w\models_{\mathrm{LTLf}}\varphi)\mathbin{\oplus}
(w\models_{\mathrm{LTLf}}\psi)\}$.
Therefore, an accepting word is exactly a distinguishing word, and the
emptiness test returns $\mathrm{false}$ exactly when the formulas are equivalent over
finite words.

\emph{Progression.} Algorithm~\ref{alg:distinguish-prog-app} in
Appendix~\ref{app:progression} is the search of
Section~\ref{sec:distinguish-prog}. By
Lemma~\ref{lem:distinguish-prog-app}, it terminates, returns a word if
and only if a distinguishing word exists, and returns the first
distinguishing word in length-lexicographic order, which is a shortest
one.
\end{proof}

\begin{proof}[Proof of Proposition~\ref{prop:sample-consistency}]
Candidate generation filters by consistency, and selection is restricted to
the filtered set. The invariant is restored after every new answer.
\end{proof}

\begin{proof}[Proof of Theorem~\ref{thm:bounded-identification}]
Every label is given by $\Target$, so the equivalence class of $\Target$
is never removed from the version space. Each query separates $\Hyp$ from
a rival, so the formulas equivalent to one of them become inconsistent:
every query removes at least one class, and removing an equivalent rival
needs no query. Hence at most $|\mathcal H/{\equiv}|-1$ refinement queries are asked.
Under the exhaustive rule, refinement on $\mathcal H$ ends only when every
remaining formula is equivalent to $\Hyp$; since the class of $\Target$
remains, $\Hyp\equiv\Target$.
\end{proof}

\begin{proof}[Proof of Theorem~\ref{thm:env-identification}]
As in the proof of Theorem~\ref{thm:bounded-identification}, with
$\equiv_{\mathcal E}$ in place of $\equiv$: formulas in one class of
$\equiv_{\mathcal E}$ receive the same label on every feasible query, the
class of $\Target$ is never removed, and each feasible witness removes at
least one class, so at most $N_{\mathcal E}-1$ refinement queries are asked. At
termination no remaining formula can be distinguished from $\Hyp$ by a
feasible trace, so $\Hyp\equiv_{\mathcal E}\Target$.
\end{proof}

\begin{proof}[Proof of Theorem~\ref{thm:pac}]
Let $\epsilon=\ln(1/\delta)/P$. If
$\operatorname{err}_{\mathcal D}(\Hyp,\Target)>\epsilon$, all $P$
independent tests agree with $\Hyp$ with probability at most
$(1-\epsilon)^P\le e^{-\epsilon P}=\delta$. Hence, with probability at
least $1-\delta$, some test disagrees or the error is at most $\epsilon$.
\end{proof}

\section{Benchmark environments}
\label{app:environments}
The environment experiments use four configurations from three benchmark
families: Office world, Taxi, and MiniGrid DoorKey. Table~\ref{tab:environment-summary}
lists their reachable state counts and the number of target formulas.
The target formulas are given in Table~\ref{tab:benchmarks}.

\begin{table}[H]
\centering
\small
\caption{Environment configurations used in the experiments.}
\label{tab:environment-summary}
\begin{tabular}{@{}lrr@{}}
\toprule
\textbf{Environment} & \textbf{States} & \textbf{Targets} \\
\midrule
Office world & 108 & 7 \\
Taxi & 101 & 3 \\
DoorKey $5\times5$ & 62 & 4 \\
DoorKey $6\times6$ & 154 & 2 \\
\bottomrule
\end{tabular}
\end{table}

\paragraph{Common representation.}
Each environment is represented as a finite deterministic labelled
transition system with a fixed start state. A trace records the label of
the initial state followed by the label after each action. Labels are
projected onto the propositions used by the target formula. Random
rollouts choose actions uniformly. Successful delivery in Taxi and goal
states in DoorKey are made absorbing: subsequent actions leave the state
and its label unchanged.

\paragraph{Office world.}
The Office world~\cite{toroicarte2022reward} is a $12\times9$ grid divided
into $3\times3$ rooms connected by doorways. The agent starts at $(2,1)$
and moves up, right, down, or left; a move blocked by a wall leaves it in
place. The propositions mark coffee ($f$), mail ($e$), the office ($g$),
plants ($n$), and four designated locations ($a,b,c,d$). The seven targets
use the proposition sets $\{f,g,n\}$, $\{e,g,n\}$, or $\{a,b,c\}$ and
specify avoidance or ordered visits to labelled locations.

\begin{figure}[H]
\centering
\includegraphics[width=0.6\linewidth]{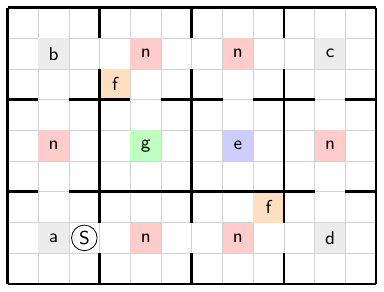}
\caption{Office world, drawn from the implemented map. Thick lines denote
walls and gaps denote doorways. $S$ marks the start; the other labels
identify coffee ($f$), mail ($e$), the office ($g$), plants ($n$), and
locations $a,b,c,d$.}
\label{fig:office-world}
\end{figure}

\paragraph{Taxi.}
The Taxi benchmark~\cite{dietterich2000hierarchical} uses Gymnasium's
\texttt{Taxi-v4}~\cite{towers2025gymnasium}, with rain and passenger
relocation disabled to obtain deterministic transitions. Reset seed $0$
fixes the taxi's initial position, passenger location, and destination.
The six actions are south, north, east, west, pickup, and dropoff.
Proposition $p$ holds while the passenger is in the taxi, and $g$ holds
once the passenger has been delivered. The three targets specify delivery,
pickup followed by delivery, or a response requirement for carrying a
passenger.

\begin{figure}[H]
\centering
\includegraphics[width=0.45\linewidth]{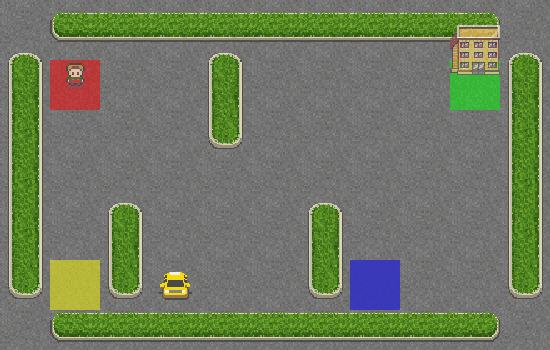}
\caption{Illustrative Taxi environment. Image: still frame from the
\href{https://gymnasium.farama.org/v1.3.0/environments/toy_text/taxi/}{official
Gymnasium documentation}. The pictured taxi and passenger positions need
not match the experiment's reset seed.}
\label{fig:taxi-environment}
\end{figure}

\paragraph{MiniGrid DoorKey.}
The DoorKey benchmarks~\cite{chevalierboisvert2023minigrid} use
\texttt{MiniGrid-DoorKey-5x5-v0} and
\texttt{MiniGrid-DoorKey-6x6-v0}, both with reset seed $0$. States record
the agent's position and orientation, the carried object, and the grid
configuration. The available actions are turn left, turn right, move
forward, pickup, and toggle; drop and done are omitted, and the original
episode step limit is ignored. Proposition $k$ denotes carrying the key,
$o$ an open door, and $g$ reaching the goal. The $5\times5$ layout has
four targets and the $6\times6$ layout has two, covering goal reachability
and requirements involving the key and door.

\begin{figure}[H]
\centering
\begin{minipage}[t]{0.3\linewidth}
\centering
\includegraphics[width=\linewidth]{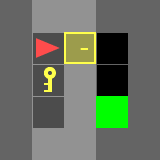}\\
\small (a) DoorKey $5\times5$
\end{minipage}\qquad
\begin{minipage}[t]{0.3\linewidth}
\centering
\includegraphics[width=\linewidth]{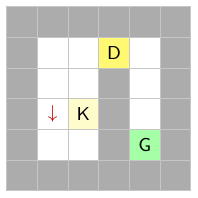}\\
\small (b) DoorKey $6\times6$
\end{minipage}
\caption{DoorKey environments. (a) Illustrative $5\times5$ layout from the
\href{https://minigrid.farama.org/environments/minigrid/DoorKeyEnv/}{official
MiniGrid documentation}; its initial configuration may differ from the
experiment's. (b) Grid drawn from the $6\times6$ experiment with reset
seed $0$: $K$ denotes the key, $D$ the locked door, $G$ the goal, and the
arrow the agent's initial position and orientation. In (a), the key and
door are yellow, the goal is green, and the agent is a red triangle.}
\label{fig:doorkey-environments}
\end{figure}

\end{document}